\documentclass{PoS}
  
\title{The Nuclear Structure of 3C\,84 with Space VLBI (RadioAstron) 
Observations}

\ShortTitle{RadioAstron Observations of 3C\,84}

\author{
\speaker{                                                                      
Gabriele Giovannini}$^{1,2}$,
Monica Orienti$^2$,
Tuomas Savolainen$^{3,4}$,
Hiroshi Nagai$^{5}$,
Marcello Giroletti$^{2}$,
Kazuhiro Hada$^{2,5}$,
Gabriele Bruni$^{4}$,
Jeffrey Hodgson$^{4}$,
Mareki Honma$^{5}$,
Motoki Kino$^{6}$,
Yuri Y. Kovalev$^{7}$,
Thomas Krichbaum$^{4}$,
Sang-Sung Lee$^{6}$,
Andrei Lobanov$^{4}$,
Bong Won Sohn$^{6}$,
Kirill Sokolovsky$^{7}$,
Peter Voitsik$^{7}$,
J.~Anton Zensus$^{4}$.

\\
$^1$Dipartimento di Fisica e Astronomia, Bologna University, Italy;
$^2$Istituto di Radioastronomia/INAF, Italy;
$^3$Aalto University Mets\"ahovi Radio Observatory, Finland.
$^4$Max-Planck-Institut f\"ur Radioastronomie, Germany;
$^5$National Astronomical Observatory of Japan, Japan;
$^6$Korea Astronomy and Space Science Institute, Republic of Korea;
$^7$Astro Space Center of Lebedev Physical Institute, Russia;

\\
E-mail: \email{ggiovann@ira.inaf.it}
}

\abstract{The radio galaxy 3C\,84 is a representative of
  gamma-ray-bright misaligned active galactic nuclei (AGN) and one of
  the best laboratories to study the radio properties of subparsec
  scale jets. We discuss here the past and present activity of the
  nuclear region within the central 1\,pc and the properties of
  subparsec-sized components C1, C2 and C3. We compare these results
  with the high resolution space-VLBI image at 5\,GHz obtained with
  the RadioAstron satellite and we shortly discuss the possible
  correlation of radio emission with the gamma-ray emission. }

\FullConference{12th European VLBI Network Symposium and Users Meeting,\\
		7-10 October 2014\\
		Cagliari, Italy}

\begin{document}

\section{Introduction}
The radio source 3C\,84 is associated with the giant galaxy NGC\,1275
(z = 0.0176), which is the Brightest Cluster Galaxy (BCG) of the
relaxed cooling flow cluster Abell 426 in Perseus. 3C\,84 is known to
be a bright radio source and has been studied extensively since the
early days of radio astronomy. At low resolution the radio emission is
extended and diffuse with a halo-core structure typical of BCG
galaxies at the center of cooling clusters.  In the central 10-pc
scale region, there is a pair of symmetric lobes, with evidence of
free-free absorption for the Northern one (Walker et al. 2000). These
lobes were probably formed by the jet activity originating in the 1959
outburst (e.g., Asada et al. 2006). At higher angular resolution the
bright central region is resolved into a bright core and a one-sided
complex jet emission. According to Liuzzo et al. (2010), the one-sided
emission is likely due to Doppler boosting effects of a highly
relativistic jet at a small angle to our line-of-sight. The jet
velocity decreases quickly because of the strong interaction between
the jet and the cooling gas, giving rise to the symmetric structure
visible in the central 10\,pc region. The radio flux has been
monitored since 1960, and episodes of violent flux increase have been
reported (Kellermann and Pauliny-Toth 1968; O'Dea et al. 1984).  In
the mid-1980s, the radio flux became exceptionally bright, more than
60\,Jy at centimetre wavelengths, and then it subsequently decreased
to $\sim$10 Jy by the early 2000s (Nagai et al. 2012).  Around 2005,
it was reported that the radio flux started to increase again (Abdo et
al. 2009). Very Long Baseline Interferometry (VLBI) observations
revealed that this flux density increase originated in the central
pc-scale region and it was accompanied by an ejection of a new jet
component (see the next Section).

NGC\,1275 was detected in $\gamma$-rays by the 
{\it Fermi} Large Area Telescope (LAT) in
August 2008 (Abdo et al. 2009). This detection is particularly
noteworthy, because the source was not detected by the 
Energetic Gamma Ray Experiment Telescope (EGRET)
of the Compton Gamma Ray Observatory (GRO) (Reimer et al. 2003). 
The new $\gamma$-ray flux density is about seven
times higher than the upper limit from the EGRET
observations. Intriguingly, the higher $\gamma$-ray emission coincides
with the increase of the radio flux density.

In the next sections we will discuss the sub-parsec structure of this
source and present a new high-resolution space-VLBI image obtained
with RadioAstron. The redshift of 3C\,84 gives an angular-to-linear
size conversion factor of 0.344 pc/mas, assuming H$_0$ = 71 km
s$^{-1}$ Mpc$^{-1}$, $\Omega_M$ = 0.27 and $\Omega_{\Lambda}$ =
0.73. We assume a jet orientation with respect to the viewing angle to
be 20$^\circ$--25$^\circ$ (Abdo et al. 2009, Nagai et al. 2014,
Tavecchio and Ghisellini 2014).

\section{Sub-parsec Radio Structure}

High resolution 43 and 86\,GHz VLBI images taken at the epochs before
$\sim$ 2005 (e.g. Dhawan et al. 1998), show a core (C1) and a nearby
bright knot separated by 0.5\,mas at a position angle
$\sim160^\circ$. Beyond 0.5\,mas, there is an abrupt bend in the jet
toward a position angle of 235$^\circ$, where a diffuse component (C2
here, D in Dhawan et al. 1998) is present at a distance of
1.2\,mas. In the first 4 epochs (from 2006/134 to 2007/258) of the
multi-epoch monitoring presented by Nagai et al. (2010), the source
structure is the same and similar to that seen in the $\gamma$-ray
quiet phase. A clear change in the source structure is evident after
the 2007/258 image and a new component C3 has emerged from the central
core.

Using the VLBA archive data and monitoring observations with the VLBI
Exploration of Radio Astrometry (VERA) array, Suzuki et al. (2012)
found that this new component was ejected from the core after 2003
November 20th. The component C3 is advancing with an increasing
apparent velocity (0.1c in November 2003, 0.47c in November 2008)
toward a position angle of $\sim$ 170$^\circ$. This direction is
clearly unrelated to that of the component C2. Moreover, the jet PA
and structure are quite different from the previous images. The flux
density of C3 and C1 have clearly increased after 2008 and at 43\,GHz
they showed about the same flux density (see Fig.~7 in Suzuki et
al. 2012).  After 2009 the core flux density at 15\,GHz has been
almost constant while the flux density of C3 showed a further
increase. The core flux density was 4.84\,Jy and the flux density of
C3 was 9.85\,Jy in the MOJAVE 15\,GHz image on February 11, 2011 (see
Lister et al. 2009 for the details of the MOJAVE project).

As discussed in Nagai et al. (2014), the relatively slow apparent
velocity of C3 is at odds with the one-sidedness of the radio
structure, the spectral energy distribution fit to the observed
broadband spectrum, and with the one-zone synchrotron-self-Compton
model. Therefore, we assume that the jet bulk motion is highly
relativistic, while the observed slow proper motion is related to the
jet advancing in a high density medium. In this interpretation C3 is a
hot-spot like structure due to the interaction of the new jet
structure with the external medium.

\begin{figure}
\centering
\includegraphics[width=.50\textwidth]{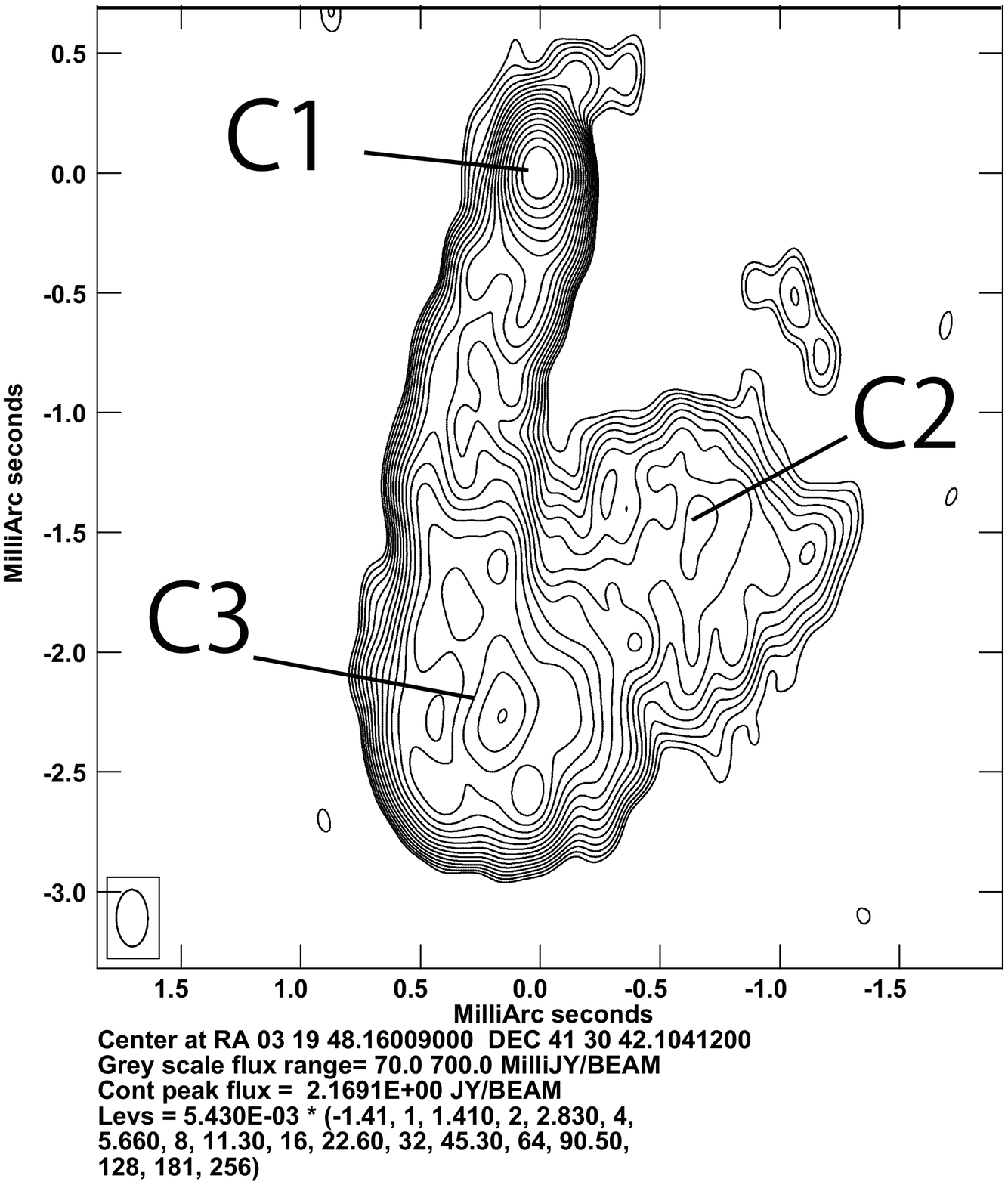}
\includegraphics[width=.48\textwidth]{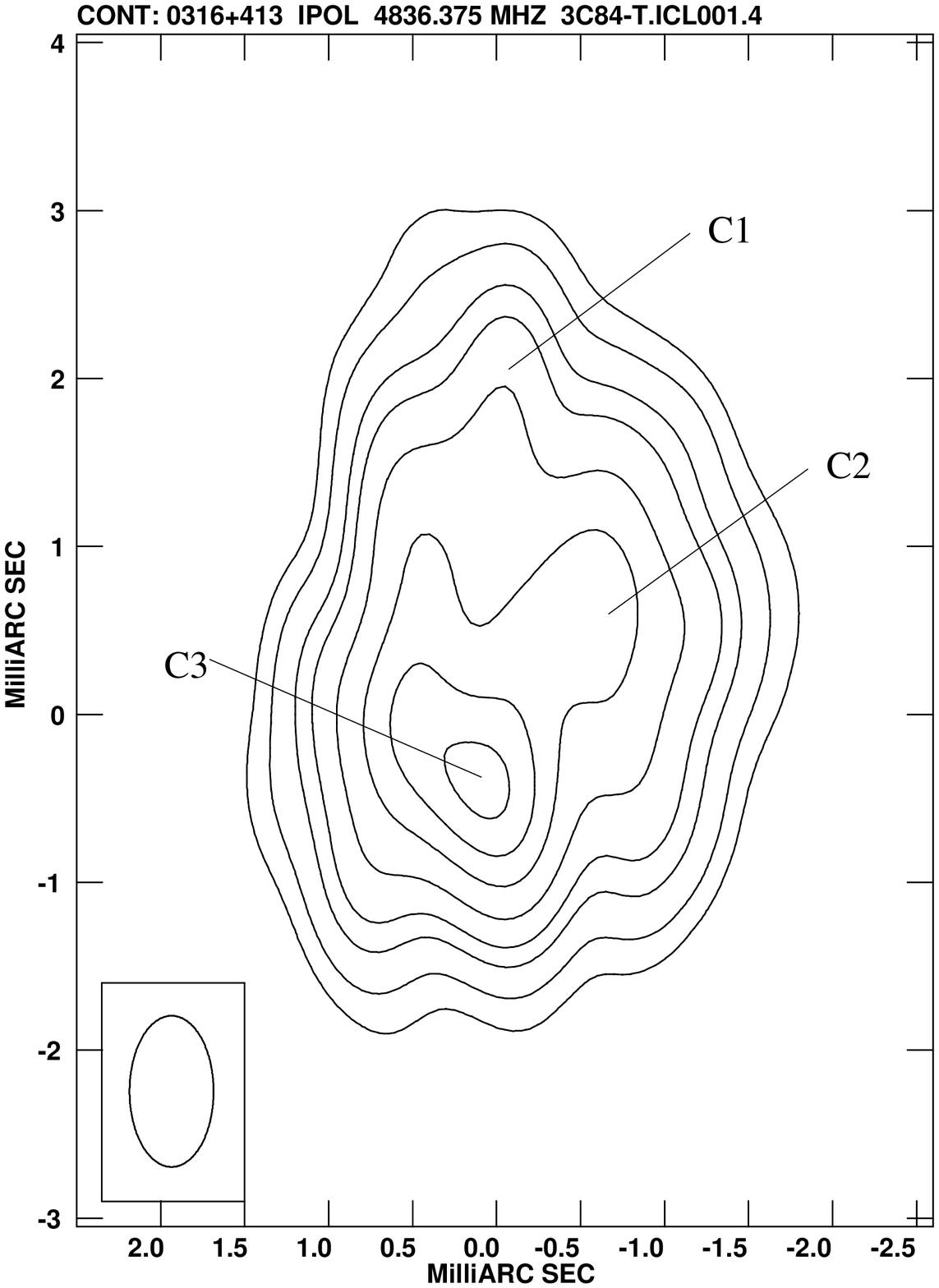}
\caption{{\it Left:} 43\,GHz total intensity map of 3C\,84 obtained
  with the VLBA in January 2013. The HPBW is 0.24 $\times$ 0.13\,mas
  in the PA of 0.7$^\circ$. For more details see Nagai et
  al. (2014). {\it Right:} Space-VLBI image of 3C\,84 at 5\,GHz. The
  HPBW is 0.9 $\times$ 0.5 mas. Countours are: 50, 100, 200, 300, 500,
  800, 1100, and 1400 mJy/beam.}
\label{f1}
\end{figure}

At 43\,GHz, the observations carried out using the VLBA on January 2013
(Nagai et al. 2014, Fig.~1) and by the Boston University's blazar
monitoring
program\footnote{http://www.bu.edu/blazars/VLBAproject.html} show an
unresolved core (C1), a complex structure coincident with C3 and a
diffuse emitting region (C2). The jet connected to C3 is
limb-brightened.  We note that in the previous images (e.g., Dhawan et
al. 1998, Romney et al. 1995, and Lister 2001) the jet activity was
connected to the component C2, oriented at a different position
angle and the jet profile was centrally peaked.

\section{Comparison of Radio and High Energy Emission}

Two $\gamma$-ray flaring periods have been reported: the first one in
April--May 2009 (Kataoka et al. 2010) and the second one in
June--August 2010 (Brown \& Adams 2011). During both of the
$\gamma$-ray flares, components C1 and C3 showed a large change of the
flux density, but on a time scale longer than that of the $\gamma$-ray
activity.  After 2010 the 15\,GHz flux density of C3 started to
increase faster than that of C1, and at present C3 is the brightest
sub-parsec feature at 15\,GHz.

\section{RadioAstron Results}

3C\,84 was observed with the 10-m Space Radio Telescope (SRT) of the
RadioAstron Mission (Kardashev et al. 2012; 2013) at 5 and 22\,GHz as
a part of the RadioAstron Nearby AGN Key Science Program (Savolainen
et al., these proceedings). Observations were carried out on September
21, 2013, outside of the regular EVN sessions because of the SRT
constrains. Science data from the SRT were transmitted in real-time to
the telemetry stations in Puschino, Russia and Green Bank, USA, where
they were recorded. The data presented here were correlated by using
the space-VLBI version of the DiFX software (dra-DiFX, see Bruni et
al. 2015) at the Bonn Astro/Geo correlator (MPIfR). This DiFX version
allows to include a space-based antenna, given the reconstructed
orbit, which for RadioAstron is provided by the ballistics group of
the Keldysh Institute of Applied Mathematics of the Russian Academy of
Sciences. Solutions for delay and delay-rate were searched for the SRT
for all the scans. When no fringes were found, typically on the
longest space baselines, delay and rate values extrapolated from
previous scans were adopted in order to optimize the correlation
window. RadioAstron fringes were clearly detected in postprocessing by
both PIMA (Petrov et al. 2011) and AIPS softwares: the shortest
baseline detection was between the SRT and Kalyazin (Kl) at $\sim$ 0.2
Earth Diameters (ED), and the longest baseline with a clear detection
(8.9 $\sigma$) was between Effelsberg (Eb) and the SRT, $\sim$ 6.9\,ED.

The ground array at 5\,GHz was: VLBA, 100-m Green Bank Telescope,
Jodrell Bank, Onsala, Shanghai, Noto, Effelsberg, WSRT,
Hartebeesthoek, and Kalyazin.  We present here preliminary results at
5\,GHz, while the 22\,GHz part of the experiment is still being
processed.

The Jansky Very Large Array (VLA) observed as a phased array (Y27)
mainly at 22\,GHz, but during the times when RadioAstron was cooling
its communications antenna, we used frequency agile ground telescopes,
including Y27, to observe 3C\,84 also at 6, 2 and 0.7\,cm. The
observations with the phased VLA also provide the standard WIDAR
correlation products and these were calibrated and imaged as well in
order to obtain accurate total flux measurements. The array was in the
C/B configuration, and we measured the flux density of the unresolved
core, obtaining 17.5\,Jy at 5\,GHz, 35.9\,Jy at 15\,GHz, 29.8\,Jy at
22\,GHz, and 29.3\,Jy at 43\,GHz. The radio spectrum is clearly
absorbed at low frequencies with a peak around 15\,GHz.

In the space-VLBI array the $(u,v)$ coverage is mostly in a narrow
range of position angles around $\sim$ 100$^\circ$. The poor, almost
one-dimensional $(u,v)$ coverage on the space baselines makes it
difficult to obtain images at full resolution, and modelling will be
required in order to utilize the data on the longest baselines.

The VLBI data calibration mostly followed the standard procedures,
except for a complicated bandbass calibration and scan-by-scan fringe
fit with solution intervals varying as a function of expected residual
acceleration of the SRT (see Savolainen et al., these proceedings).
After a detailed editing of the data, we made images limiting the
$(u,v)$ range to about 1ED ($\sim$1.0 $\times$ 10$^5$ k$\lambda$) and
we self-calibrated the ground telescope gains in phase and
amplitude. When a good image was derived we started to increase the
$(u,v)$ range in imaging and self-calibration up to the maximum
$(u,v)$ range $\sim$ 10$^6$ k$\lambda$. We used a super-uniform
weighting, which gives a low weight to short baselines, and we derived
images at different resolutions using a combination of the $(u,v)$
range and weight.

In Fig.~1, we present a space-VLBI image with a HPBW of $0.9 \times
0.5$\,mas (PA = 0$^\circ$). In the image the nuclear component is weak
due to (self-)absorbtion.  The jet from the core to the component C3 is
not resolved transversally. Component C3 is the brightest component
within the central substructure and it is compact at 5\,GHz. Component
C2 shows a diffuse morphology.


We compared the VLBA data at 43\,GHz (at an epoch close in time,
Aug. 2013) to RadioAstron data at 5\,GHz by making images with the
same cell size and angular resolution and matching as much as possible
the $(u,v)$ coverage. In the nuclear region and in the beginning of
the jet, we have a spectral index $\alpha$ $\sim$ -1.4 (S($\nu$)
$\propto$ $\nu^{-\alpha}$), i.e. these parts are strongly
(self-)absorbed. The C3 component has a spectral index of $\alpha$$\sim$
-0.7, also indicating self-absorption and thus confirming its
compactness. The jet and C2 region show $\alpha$ $\sim$ 0.2 and 1.1
respectively.

\section{Conclusions} 

From a comparison of the space-VLBI image of 3C\,84 at 5\,GHz to the
available ground-based images, we can derive some preliminary results,
while we are waiting for the higher resolution space-VLBI data at
22\,GHz:

\noindent
1) The core (C1) is very compact and (self-)absorbed between 5 and
22\,GHz.

\noindent
2) The one-sided limb-brightened jet has on average a flat radio
spectrum ($\sim$ 0.2).

\noindent
3) The C3 component at the end of the jet, is compact and
self-absorbed. At 15\,GHz it is the brightest component, and its
brightness is still increasing with time. New space-VLBI data at
higher resolution will give more information on this structure.

\noindent
4) The C2 component is resolved with a steep spectrum confirming the
interpretation that it is related to the previous activity, which has
stopped well before the current multi-wavelength activity.

\noindent
5) Properties of the previous ($\gamma$-silent) activity period
(related to component C2) and the present activity are quite different
in the jet structure, its PA, and in the high energy activity.

The lack of a strong correlation between the $\gamma$-ray flares and
the radio activity suggests that the origin of the high energy
emission may be due to a 'spine--layer' scenario discussed by Tavecchio
and Ghisellini (2014).  However, the activity of the component C3,
with an increasing flux density and a strong interaction with the
external medium, could be another source of $\gamma$-ray emission.

\vskip 0.5truecm
\noindent
{\bf Acknowledgements}
\noindent
The RadioAstron project is led by the Astro Space Center of the
Lebedev Physical Institute of the Russian Academy of Sciences and the
Lavochkin Scientific and Production Association under a contract with
the Russian Federal Space Agency, in collaboration with partner
organizations in Russia and other countries.  The National Radio
Astronomy Observatory is a facility of the National Science Foundation
operated under cooperative agreement by Associated Universities,
Inc. The European VLBI Network is a joint facility of European,
Chinese, South African and other radio astronomy institutes funded by
their national research councils. This research is based on
observations correlated at the Bonn Correlator, jointly operated by
the Max Planck Institute f\"ur Radioastronomie (MPIfR), and the
Federal Agency for Cartography and Geodesy (BKG). This research has
made use of data from the MOJAVE database that is maintained by the
MOJAVE team.  TS was partly supported by the Academy of Finland
project 274477.

\vskip 0.5truecm
\noindent
{\bf References}

\noindent
Abdo, A.A., Ackermann, M., Ajello, M., et al.: 2009, ApJ, 699, 31

\noindent
Asada, K., Kameno, s., Shen, Z.-Q., et al. 2006, PASJ, 58, 261

\noindent
Brown, A.M., Adams, J.: 2011, MNRAS, 413, 2785

\noindent
Bruni, G., et al.: 2015, PoS(EVN 2014)119

\noindent
Dhawan, V., Kellermann, K.I., Romney, J.D.: 1998, ApJL 498, L111

\noindent
Kardashev, N.S., Kovalev, Y.Y., Kellermann, K.I.: 2012, The Radio Sc Bull 
343, 22 (arXiv:1303.5200)

\noindent
Kardashev, N.S., Khartov, V.V., Abramov, V.V., et al.: 2013, ARep, 57, 153
(arXiv:1303.5013) 

\noindent
Kataoka, J., Stawarz, L., Cheung, C.C., et al., 2010, ApJ, 715, 554

\noindent
Kellermann, K.I., and Pauliny-Toth, I.: 1968, AJ, 73, 874 

\noindent
Lister, M.L.: 2001, ApJ, 562, 208

\noindent
Lister, M.L., Aller, H.D., Aller, M.F., et al.: 2009, AJ, 137, 3718

\noindent
Liuzzo, E., Giovannini, G., Giroletti, M., Taylor, G.B.: 2010, A\&A, 516, 1

\noindent
Nagai, H., Suzuki, K, Asada, K., et al.: 2010, PASJ 62, L11

\noindent
Nagai, H., Orienti, M., Kino, M., et al.: 2012, MNRAS, 423, L122

\noindent
Nagai, H., Haga, T., Giovannini, G., et al.: 2014, ApJ, 785, 53

\noindent
O'Dea, C.P., Dent, W.A., Balonek: 1984, ApJ, 278, 89 

\noindent
Petrov, L., Kovalev, Y.Y., Fomalont, E.B., Gordon, D.: 2011, AJ, 142, 35

\noindent
Reimer, O., Pohl, M., Sreekumar, P., Mattox, J.R.: 2003 ApJ, 588, 155

\noindent
Romney, J.D., Benson, J.M., Dhawan, V., et al.: 1995, PNAS, 92, 11360

\noindent
Suzuki, K., Nagai, H., Kino, M., et al.: 2012, ApJ, 746, 140

\noindent
Tavecchio, F., Ghisellini, G.: 2014, MNRAS, 443, 1224

\noindent
Walker, R. C., Dhawan, V., Romney, J. D., et al. 2000, ApJ 530, 233

%

\end{document}